\documentclass{iaus}
\usepackage{graphicx}

\title[~~Ferromagnetic properties of W bosons condensate] %% give here short title %%
{Ferromagnetic properties \\ of charged vector bosons condensate \\ in
the early universe}

\author[Gabriella Piccinelli]   
{Gabriella Piccinelli}

\affiliation{Centro Tecnol\'ogico, FES Arag\'on, Unversidad Nacional
  Aut\'onoma de M\'exico \\ Av. Rancho Seco S/N, Bosques de Arag\'on, 
Nezahualc\'oyotl \\ Estado de M\'exico 57130, Mexico
\\ email: {\tt itzamna@unam.mx} \\[\affilskip]}

\pubyear{2011}
\volume{274}  %% insert here IAU Symposium No.
\pagerange{001--004}
\jname{Advances in Plasma Astrophysics}
\editors{A. Bonanno, E. de Gouveia Dal Pino \& A. Kosovichev, eds.}
\begin{document}

\maketitle

\begin{abstract}
Bose-Einstein condensation in the early universe is considered. The
magnetic properties of a condensate of charged vector bosons are
studied, showing that a ferromagnetic state is formed. 
As a consequence, the primeval plasma may be spontaneously magnetized
inside macroscopically large domains and primordial magnetic fields
can be generated.
\end{abstract}

\firstsection % if your document starts with a section,
\section{Introduction}

{\underline{\it Cosmological Magnetic fields}}. Magnetic fields seem
to be pervading the entire universe: they have
been observed in galaxies, clusters and high
redshift objects (for observational reviews see, e.g., 
\cite[Kronberg 1994]{kronberg94} and 
\cite[Carilli \& Taylor 2002]{Carilli02}). 
Although at present there is no conclusive
evidence about their origin, their existence in the early universe
cannot certainly be ruled out. They use to be appreciated in cosmology
since they help to solve some problems, in particular in the baryogenesis
process (e.g., \cite[S\'anchez, Ayala \& Piccinelli 2007] {Angel07} and
references therein and \cite[Piccinelli \& Ayala 2004] {lectnotes},
for a review).

The determination of the physical processes able to generate magnetic
fields represents a long standing cosmological problem. The general 
approach is to
identify mechanisms for the generation of seed fields which can later
be amplified into fields on larger scales, but nowadays no completely
successful mechanism has been proposed. 

{\underline{\it Bose-Einstein condensates (BEC)}}. 
Another interesting 
phenomenon that may be present in the early
Universe is Bose-Einstein condensation: a quantum phenomenon of 
accumulation of identical bosons in the same state, which is their
lowest energy (zero momentum) state. Under 
these conditions they behave as a single macroscopic entity described
by a coherent wave function rather than a collection of separate
independent particles. 
Even though Bose-Einstein condensation had been foreseen long time 
ago (1925), it took seventy years to make the first experimental
observation, which was performed in a dilute gas of 
rubidium (\cite[Anderson el al. 1995] {Anderson:1995gf}). 
Difficulties in performing this observation were created by the
extremal conditions necessary for the condensation. Indeed,
the Bose-Einstein condensation takes place when the inter-particle
separation is smaller than their de Broglie wavelength, 
$\lambda_{dB} \sim 2\pi/\sqrt{2m T}$, so the system must be cooled
down to a very low temperature at ordinary densities. In the early
universe, the existence of condensates depends
on an interplay between density and temperature (see \cite[Dolgov,
 Lepidi \& Piccinelli 2010] {Dolgov3},
for a discussion on the conditions for which condensation 
could take place). 

In the recent years the study of BEC became an active area
of research in different fields of physics from plasma and 
statistical physics (for a review see \cite[Pethick \& Smith
  2002] {Pethick_book} and references therein) to astrophysics and
cosmology. For instance, neutral scalar BEC have been proposed as 
dark matter (\cite[Ji \& Sin 1994]{Ji94}, \cite[Matos \& Guzm\'an
  1999]{Matos99}).

The presence of charged BEC has interesting consequences in gauge 
field theories. For instance, in \cite[Dolgov, Lepidi \& Piccinelli
  (2009)] {Dolgov1}, 
electrodynamics of charged fermions and condensed scalar bosons was
considered. The screening of impurities in such plasma was found to be 
essentially different from the case when the condensate is absent.
A similar problem was considered also in the framework of an effective 
field theory (\cite[Gabadadze \& Rosen (2008), (2009), (2010)] 
{Gabadadze}),
analyzing the thermodynamical properties of the system and
focusing on the astrophysics of helium white dwarfs. 
The possible condensation of helium nuclei, previous to
crystallization, would affect the cooling process and leave
observational signatures.

The condensation of gauge bosons of weak interactions was considered
in the pioneering papers by \cite[Linde (1976), (1979)] {Linde}, where it 
was argued that, at
sufficiently high leptonic chemical potential, a classical
$W_j$ boson field could be created and, under certain 
conditions, there could be no symmetry restoration in the early universe. 

As we have seen in the previous applications, the
condensate 
can be made of either scalar or vector bosons. In both
cases, bosons are in the lowest energy state, but in the vector case
they have an additional degree of freedom: spin. The state of rest and 
identicalness inherent to the condensed particles suggests that they
could present an ordered configuration of their spin magnetic
moments. 

In this work we consider the BEC of charged $W$-bosons, which may be
formed if the cosmological lepton asymmetry happened to
be sufficiently high, i.e. if the chemical potential of neutrinos was
larger than the $W$ boson mass at this temperature, and study its
magnetic properties. We concentrate on the condensation below
electroweak phase transition,
although the phenomenon could also be considered above it. 

\section{The model}

We consider a simple example of electrically neutral plasma made of
fermions (electrons and neutrinos), gauge and scalar bosons, with zero 
baryonic number density but with a high leptonic one. This implies a 
(large) chemical 
potential for leptons and W bosons and, consequently, the
formation of a charged vector BEC. 
For simplicity we work with only one family of leptons, but this does
not influence the essential features of the result. 
Quarks may be essential for the condition of vanishing of all gauge
charge densities in plasma and for the related cancellation of the
axial anomaly but we work in the lowest order of the perturbation 
theory, where the anomaly is absent.

A caveat for a large lepton asymmetry arises from the
big bang nucleosynthesis (BBN) with strongly mixed neutrinos. It is 
shown in \cite[Dolgov et al. (2002)]{bbn-osc} that leptonic chemical 
potentials of all neutrino 
flavors are restricted by $|\mu_\nu /T |< 0.07$ at the BBN epoch. 
However, it should be noted that the entropy release from the
electroweak epoch down to the BBN epoch diminishes the lepton asymmetry, 
by the ratio of the particle species present in the 
cosmological plasma at these two epochs, which is approximately $10$. 
This helps to alleviate the bound and to allow favorable conditions for 
condensation in the early
universe, still compatible with BBN (see \cite[Dolgov,
  Lepidi \& Piccinelli 2010] {Dolgov3}, for a detailed 
discussion on the different possibilities).

The essential reactions are the direct and inverse decays of $W$: 
$W^+ \leftrightarrow e^+ + \nu\,.$
The equilibrium with respect to these processes imposes the equality
between the chemical potentials:
$\mu_W = \mu_\nu - \mu_e\,$,
and the condition of electroneutrality reads:
$n_{W^+} - n_{W^-} - n_{e^-} + n_{e{^+}} = 0$.

Considering the Lagrangian of the minimal electroweak model, we derive
the equations of motion that allow to take into account the spin-spin 
interactions of the vector bosons.

\section{Spin-spin interactions of the condensed bosons}

The spins of the individual vector bosons can be either aligned or
anti-aligned, forming, respectively, ferromagnetic and 
anti-ferromagnetic states, see, e.g., \cite[Pethick \& Smith
  (2002)] {Pethick_book}. 
The realization of one or the other state
is determined by the spin-spin interaction between the bosons. 
In the lowest angular momentum state, $l=0$, a pair of bosons may have either 
spin 0 or 2. Depending on the sign of the spin-spin coupling, one of
those states would be energetically more favorable and would be
realized at the condensation.

In solid state physics, the dominant contribution
to the spin-spin interactions comes from quantum exchange effects;
for bosons these forces are not essential and the spin-spin
interaction is determined by the electromagnetic interaction between
their magnetic moments and by their self-interactions (\cite[Dolgov,
  Lepidi \& Piccinelli 2010] {Dolgov3}). 

{\underline{\it Electromagnetic interactions}} can be found from the
analogue of the Breit equation for electrons, which leads to the 
spin-spin potential: 

	\begin{eqnarray*}
	\label{U_spin}
	U^{spin}_{em} (r)=  \frac{e^2}{4\pi m_W^2} 
%	\left( \mathbf{V}_1 \cdot \mathbf{V}_2^*\right)
	\left[% \frac{1}{\pi} 
	\frac{\left(\mathbf{S}_1 \cdot \mathbf{S}_2 \right)}{r^3}
	- 3 \, % \frac{3}{\pi} 
	\frac{\left(\mathbf{S}_1 \cdot \mathbf{r} \right)  
\left(\mathbf{S}_2
 \cdot \mathbf{r} \right)}{r^5}
	- \frac{8 \pi}{3} \left(\mathbf{S}_1 \cdot \mathbf{S}_2 \right) 
\delta^{(3)}(\mathbf{r})
	\right],
%	\left( \mathbf{V}_3 \cdot \mathbf{V}_4^*\right)
	\end{eqnarray*}
where $\mathbf{S} = - i \,{\bf W}^\dagger \times {\bf W}$ is the
spin operator of vector particles and $m_W$ its
 mass.

To calculate the contribution of this potential into the energy of two 
$W$-bosons we have to average it over their wave function. 
In particular, in the condensate case,
it is a S-wave function that is angle independent. Hence the
contributions of the first two terms in the previous equation  
mutually cancel out and only the third one remains, which has negative
coefficient. Thus the energy shift induced by the spin-spin
interaction is:
	\begin{eqnarray*}
	\label{En_shift}
	\delta E = \int \frac{d^3r}{V}  \, U^{spin}_{em} (r) = 
	- \frac{2 \, e^2}{3 \, V m_W^2} \left(\mathbf{S}_1 \cdot \mathbf{S}_2 \right),
	\end{eqnarray*}
where V is the normalization volume.

Since $S_{tot}^2 = (S_1+S_2)^2 = 4 + 2 S_1 S_2$, the average value of
 $S_1S_2$ is $S_1 S_2 = S_{tot}^2/2 - 2$.
For $S_{tot} = 2$ this term is $S_1 S_2 = 1 >0$, while for $S_{tot} =0$ it
is $S_1 S_2 = -2 <0$. Thus, for this interaction,
the state with maximum total spin is more favorable energetically 
and $W$-bosons should condense with aligned spins. 

{\underline{\it Quartic self-coupling of W}}. Due to its non-Abelian
character, the weak sector presents a quartic self-coupling in the Lagrangian: 
\begin{eqnarray*}
L_{4W}= - \frac{e^2}{2\sin^2\theta_W}\left[(W_\mu^\dagger W^\mu )^2 - 
W_\mu^\dagger W^{\mu \dagger} W_\nu W^\nu \right]=
\frac{e^2}{2\sin^2\theta_W} \left({\bf W}^\dagger \times{\bf W}\right)^2,
\label{spin-spin-W}
\end{eqnarray*}
where $\theta_W$ is the Weinberg angle.
It is assumed here that $\partial_\mu W^\mu =0 $ and thus only the
spatial 3-vector ${\bf W}$ is non-vanishing, while $W_t =0$.

The interaction potential is given by the Fourier transform of this
term which, with proper (nonrelativistic) normalization, leads to:
\begin{eqnarray*}
U^{(spin)}_{4W} = \frac{e^2}{8 m_W^2 \sin^2 \theta_W} \left( {\bf S }_1 
\cdot {\bf S}_2 \right) \delta^{(3)} ({\bf r})\,.
\label{U-spin-W}
\end{eqnarray*}

Thus, the contribution of the quartic self-coupling of $W$
to the spin-spin interaction has opposite sign to the electromagnetic
term and tends to favour an antiferromagnetic state. Nonetheless, in
the standard model, the electromagnetic term is dominant.

\section{Discussion and Conclusions}

{\underline{\it Generation of primordial magnetic fields.}} We have studied the
 Bose-Einstein condensation of charged weak
bosons, driving special attention to the magnetic properties
of the condensate. The spin-spin interaction has two contributions: 
the electromagnetic one - leading to a ferromagnetic state - and a
 contact quartic self-interaction that produces an antiferromagnetic state.
We have found that the former dominates, generating 
a macroscopically large ferromagnetic configuration. 
We then expect that the primeval
plasma, where such bosons  condensed, can be spontaneously
 magnetized. The typical size of the magnetic
domains is determined by the cosmological horizon at the moment of the
condensate evaporation. The latter takes place when the neutrino chemical
potential, which scales as temperature in the course of cosmological
cooling down, becomes smaller than the $W$ mass at this temperature.
Large scale magnetic fields created by the ferromagnetism of $W$-bosons
might survive after the decay of the  condensate due to the 
conservation of the magnetic flux in plasma with high electric conductivity.
Such magnetic fields at macroscopically large scales could be the seeds of the
observed larger scale galactic or intergalactic magnetic fields. 

{\underline{\it Some considerations:}}
The long range interactions between magnetic moments can, in
principle, be screened by plasma physics, while the local quartic
interaction cannot. This could change the relative strength of these
two effects. However, in the broken phase, the problem is reduced to
that of pure QED, where it is known that magnetic forces are not
screened. On the other hand, the situation is not clear in non-Abelian
theory and it could happen that, in the symmetric phase, the screening might
inhibit magnetic spin-spin interaction. 
Interactions with relativistic electrons and positrons are neglected.
In principle, they could distort the spin-spin
interactions of $W$  by their spin or orbital motion and thus destroy 
the attraction of parallel spins of $W$. However, it looks hardly
possible because electrons are predominantly ultra-relativistic 
and they cannot be attached to any single $W$ boson to counterweight 
its spin. Finally, the scattering of electrons (and quarks) on
$W$-bosons may lead to the spin flip of the latter, but in thermal 
equilibrium this process does not change the average value of the spin 
of the condensate.

\end{document}